\documentclass[10pt]{amsart}
\usepackage{latexsym,amssymb,amsmath,amscd,amsthm}
\numberwithin{equation}{section}
\theoremstyle{remark}
\newtheorem{theorem}{{\bf THEOREM}}[section]

\newtheorem{example}{{\bf EXAMPLE}}[section]
\newcommand{\bq}{\begin{equation}}
\newcommand{\bea}{\begin{array}}
\newcommand{\eea}{\end{array}}

\newcommand{\gD}{\Delta}
\newcommand{\gl}{\lambda}

\newcommand{\gb}{\beta}

\newcommand{\go}{\omega}
\newcommand{\gO}{\Omega}

\newcommand{\gag}{\gamma}

\newcommand{\pp}{\partial}

\newcommand{\na}{\nabla}

\newcommand{\bl}{\blacklozenge}
\newcommand{\bs}{\blacksquare}

\newcommand{\gS}{\Sigma}

\newcommand{{\DDD}}{D\!\!\!\!\!\!-}

\title{SOME FUNDAMENTAL ASPECTS OF A QUANTUM POTENTIAL}
\author{Robert Carroll\\University of Illinois, Urbana, IL 61801}

\date{June, 2005\thanks{email: rcarroll@math.uiuc.edu}}

\begin{document}

\bibliographystyle{plain}

\begin{abstract}
We show that given an essentially arbitrary $Q(x,t,\hbar)$ there are ``generalized"
quantum theories having Q as their quantum potential.

\end{abstract}

\maketitle


\section{INTRODUCTION}
\renewcommand{\theequation}{1.\arabic{equation}}
\setcounter{equation}{0}

Let a function $Q(x,t,\hbar)\sim Q(x,t)\sim Q$ be given (with properties to be determined).
Following \cite{c1}, in order for Q to be a quantum potential with a 
Schr\"odinger equation (SE)
$(\clubsuit)\,\,-(\hbar^2/2m)\gD\psi+V\psi=i\hbar\pp_t\psi$ (where $\psi=R(x,t)exp[iS(x,t)/
\hbar]$) one requires that $(\spadesuit)\,\,Q=-(\hbar^2/2m)(\gD R/R)$.  We ignore here
"delicate" situations where $S=constant$ etc. (cf. \cite{c1,c3,f1}).  From $(\spadesuit)$
one derives quantum Hamilton-Jacobi equations (QHJE) of the form
\bq\label{1.1}
\pp_tR^2+\frac{1}{m}\na(R^2\na S)=0;\,\,\pp_tS+\frac{1}{2m}(\na S)^2+Q+V=0
\end{equation}
($V$ will be assumed to have suitable properties as needed).  The plan here is to solve
$(\spadesuit)$ for $R=R(Q,f(t),g(t))$ and then fit this into \eqref{1.1}.  
\\[3mm]\indent
One should note a few known limitations relating quantum and classical mechanics via the
quantum potential (cf. \cite{b1}).  Thus 
\begin{enumerate}
\item
for a free particle in 1-dimension (1-D) one has
possibilities such as $\psi'=Aexp[i(px-(p^2t/2m))/\hbar]$ and $\psi''=Aexp[-i(px+(p^2t/2m))/
\hbar]$ in which case $Q=0$ for $\psi'$ and $\psi''$ separately but for $\psi=
(\psi'+\psi'')/\sqrt{2}$ there results $Q=p^2/2m$ 
($p\sim\hbar k$ here - cf. Remark 3.1).  Hence $Q=0$ depends on the wave function and
cannot be said to represent a classical limit.
\item
For $V=m\go^2x^2/2$ and a stationary SE one has solutions of the form $\psi_n(x)=c_nH_n(\xi
x)exp(-\xi^2x^2/2)$ where $\xi=(m\go\hbar)^{1/2},\,\,c_n=(\xi/\sqrt{\pi}2^nn!)$, and $H_n$ is a
Hermite function. One computes that $Q=\hbar\go[n+(1/2)]-(1/2)m\go^2x^2$.  Hence $\hbar\to 0$
does not imply
$Q\to 0$ and moreover $Q=0$ corresponds to $x=\pm\sqrt{(2\hbar/m\go)[n+(1/2)]}$ so not all
systems in quantum mechanics (QM) have a classical limit.
\end{enumerate}
Therefore evidently in general one cannot identify QM as quantization of classical systems or 
the quantum potential as a vehicle to generate QM since in particular
there are physically realizable classical situations that
cannot be reached as the limit of some QM system, $\hbar\to 0$ and $Q=0$ are generally different
concepts, and the condition
$Q=0$ can depend on the wave function.  On the other hand we have exhibited and studied in 
\cite{c1,c4,c5} a vast collection of examples and situations where the quantum potential Q
in e.g. Schr\"odinger and Klein-Gordon equations  plays
a fundamental role in connection with quantum fluctuations, diffusion, Weyl geometry, entropy,
etc.  In other words there are physical and geometrical origins of quantum potentials and 
such interaction of QM and geometry is surely related to the elusive understanding of ``quantum
gravity" (whatever that may be).

\section{THE ELLIPTIC EQUATION}
\renewcommand{\theequation}{2.\arabic{equation}}
\setcounter{equation}{0}

Suppose $R=0$ outside of some region $\gO\subset {\bf R}^3$ (since $R^2\sim |\psi|^2$ is a
probability density this would be reasonable for many QM problems).  Consider then the elliptic
equation 
\bq\label{2.1}
L(R)=-\gD R-\gb QR;\,\,\gb=\frac{2m}{\hbar^2}
\end{equation}
Given say $R\in H_0^1(\gO)$ this is associated with a bilinear form ($R_i=\pp_iR$)
\bq\label{2.2}
B(R,\phi)=\int_{\gO}\sum R_i\phi_i-\gb\int_{\gO}Q R\phi
\end{equation}
Recall $\|v\|^2_{H_0^1}=\|v\|^2_{L^2}+\sum\|v_i\|^2_{L^2}$ (cf. \cite{c2} for notation)
so $|B(R,\phi)|\leq c\|R\|_{H^1_0}\|\phi\|_{H_0^1}$ when $Q\in L^{\infty}$ for example.
Further for $\gag>\gb sup|Q|$ one has 
\bq\label{2.3}
B(R,R)+\gag\|R\|^2_{L^2}\geq c'\|R\|^2_{H_0^1}
\end{equation}
so by Lax-Milgram for example one can say that for $\mu\geq\gag$ there exists a unique solution
of $LR+\mu R=0$ (cf. \cite{c2,e1,l1}).  On the other hand if e.g. $Q\leq 0$ one has 
$\gb Q\geq 0$ and
\bq\label{2.4}
B(R,R)\geq c''\|R\|^2_{H_0^1}
\end{equation}
(note for $f\in H_0^1$ one has $\|f\|^2_{L^2}\leq \hat{c}\sum\|f_i\|^2_{L^2}$).  Consequently
(cf. \cite{e1} for proof)
\begin{theorem}
For $Q\in L^{\infty}$ and $Q\leq 0$ the equation $\gD R=\gb QR$ has a unique solution in 
$H_0^1$.  If $Q\in L^{\infty}$ and $\mu>\gb sup|Q|$ then there exists a unique solution of
$\gD R+\mu R=\gb QR$ in $H_0^1$.  Further there is an at most countable set $\gS\subset {\bf
R}$ of eigenvalues $\gl_k\to\infty$ such that $\gD R=\gb QR+\gl R$ has a unique solution
if and only if $\gl\notin\gS$.  In particular if $0\notin \gS$ then $-\gD R=\gb QR$ has a
unique solution for any $Q\in L^{\infty}$.
\end{theorem}
\indent
{\bf REMARK 2.1.}
These are typical results for $H_0^1$ and using methods of duality and functional analysis
one can produce various theorems involving solutions $u\in H^1(\gO)$ or other Sobolev spaces
(cf. \cite{c2,l1,l2,l3}).  Here one notes that $H^{-1}$ is the dual of $H_0^1$ where $f\in
H^{-1}$ means that there exist $f^i\in L^2$ such that $<f,v>=\int_{\gO}(f^0v+\sum
f^iv_{x_i})dx$ for $v\in H_0^1$.  The theorems for solutions of $\gD R+\mu R=\gb QR$ or
$(L+\mu) R=0$ above are special cases of $L_{\mu}u=g$ for say $g\in L^2$ and one can extend
easily to see that $L_{\mu}:\,\,H_0^1(\gO)\to H^{-1}(\gO)$ is an isomorphism (cf. \cite{e1}).
$\hfill\bs$

\section{THE HAMILTON JACOBI EQUATION IN 1-D}
\renewcommand{\theequation}{3.\arabic{equation}}
\setcounter{equation}{0}

The plan now is to solve for $\na S$ from the first equation in \eqref{1.1} and then reduce
the HJ equation to a simple ordinary differential equation in $t$.  This will avoid the need
of considering e.g. viscosity solutons of the HJ equation (see e.g. \cite{e1}).  Thus set
first $(\bl)\,\,\na S=p$ with $\dot{q}=p/m$ and consider
\bq\label{3.1}
m\pp_tR^2+\na(R^2p)=0
\end{equation}
with R given via Theorem 2.1 as $R(x,t)=R(Q(x,t),x)$ (no arbitrary functions of t are
introduced here and we recall that Q may depend on $\hbar$).  In 1-D this is
$m\pp_tR^2+\pp(R^2p)=0$ from which
\bq\label{3.2}
R^2p=-\int^xm\pp_tR^2dx +mf(t)
\end{equation}
($f$ ``arbitrary").
Writing the HJ equation as $\pp_tS+(1/2m)p^2+Q+V=0$ one arrives at
\bq\label{3.3}
\pp_tS=-Q-V-\frac{1}{2R^2}\left[f(t)-\int^x\pp_tR^2dx\right]^2
\end{equation}
from which
\bq\label{3.4}
S=-\int^t(Q+V)dt-\frac{1}{2}\int^t\frac{1}{R^4}\left[f(t)-\pp_t\int^xR^2dx\right]^2+g(x)
\end{equation}
for $g$ a suitable ``arbitrary" function and $\pp_tF=f$ arbitrary.
\begin{theorem}
In 1-D, given a solution $R=R(Q(x,t),x)$ of $-\gD R=\gb QR$ as in Theorem 2.1, one can find a
solution $S=S(Q(x,t),x,t)$ in the form \eqref{3.4}, where $f,\,g$ are suitable ``arbitrary"
functions and $V(x)$ is given.  This will represent a ``generalized" quantum theory in some
sense determined by Q (V being a suitable function).
\end{theorem}
\begin{example}
Ignoring temporarily any restriction $R\in H_0^1$ (which is also violated in Items 1 and 2 of
Section 1) consider $Q=0$ so that $R''=0$ implies $R=a(t)x+b(t)$ for suitable $a,\,b$.  Then
\eqref{3.2} implies (for 1-D and suitable $f(t)$)
\bq\label{3.5}
(ax+b)^2p=-m\int^x\pp_tR^2+mF_t(t)=-m\pp_t\int^x(ax+b)^2dx+m\pp_tF(t)\Rightarrow
\end{equation}
$$\Rightarrow p=\frac{-m}{(ax+b)^2}\pp_t\left[\frac{(ax+b)^3}{2a}\right]+m\pp_tF(t)$$
\end{example}
Now differentiating in $x$ one can write the second equation in \eqref{1.1} as 
($\dot{p}=\pp_tp$ and $p'=\pp_xp$)
\bq\label{3.6}
\dot{p}+\frac{1}{m}pp'+\pp V=0
\end{equation}
Consequently one obtains an expression for a putative $\pp V$ in the form
\bq\label{3.7}
\pp V=-\dot{p}-\frac{1}{m}pp'=m\pp_t\left\{\frac{1}{(ax+b)^2}\pp_t\left[\frac{(ax+b)^3}
{3a}+F\right]\right\}-
\end{equation}
$$-m\left\{\left[\frac{1}{(ax+b)^2}\pp_t\left(\frac{(ax+b)^3}{3a}+F\right)\right]
\pp_x\left[\frac{1}{(ax+b)^2}\pp_t\left(\frac{(ax+b)^3}{3a}+F\right)\right]\right\}$$
Consider now special cases $a$ or $b$ equal to zero with $F=0$.  For $a=0$ one has
\bq\label{3.8}
b^2p=-m\pp_tb^2x\Rightarrow p=-mx\pp_tlog(b^2)=-2mx\pp_tlog(b)
\end{equation}
Hence
\bq\label{3.9}
\dot{p}=-2mx\pp_t^2log(b);\,\,p'=-2m\pp_tlog(b);
\end{equation}
$$\pp V=-\dot{p}-\frac{1}{m}pp'=
2xm[\pp_t^2log(b)-2(\pp_tlog(b))^2]$$
Thus if e.g. $b=exp(\pm ct)$ with $log(b)=\pm ct$ and $\pp_tlog(b)=\pm c$ one has
$(\bullet)\,\,\pp V=-4mxc^2$ which seems like a conceivable physical potential $V\sim
-2mc^2x^2$.  If $b=0$ one gets
\bq\label{3.10}
a^2x^2p=-m\pp_t\int^xa^2x^2dx=-\frac{mx^3}{3}\pp_ta^2\Rightarrow p=-\frac{2mx}{3}\pp_tlog(a)
\end{equation}
leading to
\bq\label{3.11}
\dot{p}=-\frac{2mx}{3}\pp_t^2log(a);\,\,p'=-\frac{2m}{3}\pp_tlog(a)
\end{equation}
Hence for $a=exp(\pm ct)$ as before one has $\pp_tlog(a)=\pm c$ and 
\bq\label{3.12}
\pp V=-\frac{4mc^2x}{9}\sim V=-\frac{2}{9}mc^2x^2
\end{equation}
much as in the case $a=0$.  One notes that the potential V in both these situations
corresponds to the negative of the potential in Item 2 of Section 1.$\hfill\bs$
\\[3mm]\indent
{\bf REMARK  3.1.}
Note that the situation of Item 1 in Section 1 can also be attained here for $Q=0$.  Indeed
take $R=1$ (i.e. $a=0$ and $b=1$ in Example 3.1).  Then $F\ne 0$ with $R^2p=p=m\dot{F}(t)$ and
since $Q=V=0$ one has $\pp_tS=-(1/2)F(t)$.  Note in Item 1 p is used for $\hbar k$ where k is a
frequency and e.g. $S=\hbar kx-(\hbar^2k^2/2m)t$ with $S_t=-\hbar^2k^2/2m\sim -E$ so here
$(1/2)F(t)=E$.  Then the HJ equation becomes $-E+(1/2m)S_x^2=0$ with $S_x=\hbar k$.
Alternatively (referring to Item 1) for $\psi=(\psi'+\psi'')/\sqrt{2}$ and $Q=\hbar^2k^2/2m$
with $V=0$ we have
\bq\label{3.13}
R=\sqrt{2}ACos(kx);\,\,R''/R=-k^2;\,\,Q=\frac{k^2\hbar^2}{2m};\,\,S=-\frac{k^2\hbar^2t}{2m};
\,\,S_t=-\frac{k^2\hbar^2}{2m};\,\,S_x=0
\end{equation}
Consequently $S_t+(1/2m)(S_x)^2+Q+V=-(k^2\hbar^2/2m)+(k^2\hbar^2/2m)=0$ and one sees that
the same SE can arise from different quantum potentials.
$\hfill\bs$
\\[3mm]\indent
{\bf REMARK 3.2.}
Evidently with unique solutions R as in Theorem 2.1 one should arrive at fewer possibilities
in the construction of S.  Otherwise the map $SE\to Q$ is seen to be
possibly multivalued.
$\hfill\bs$

\end{document}